\documentclass[aps,pra,twocolumn,noshowpacs,showkeys]{revtex4-1}
\usepackage{graphicx, ulem}
\usepackage{amsmath, amssymb}
\usepackage[colorlinks=true,urlcolor=blue,citecolor=blue,linkcolor=blue]{hyperref}

\begin{document}

\title{Anomalous quantum correlations in the motion of a trapped ion}

\author{F. Krumm}\email{fabian.krumm@uni-rostock.de}\affiliation{Arbeitsgruppe Theoretische Quantenoptik, Institut f\"ur Physik, Universit\"at Rostock, D-18059 Rostock, Germany}
\author{W. Vogel}\affiliation{Arbeitsgruppe Theoretische Quantenoptik, Institut f\"ur Physik, Universit\"at Rostock, D-18059 Rostock, Germany}

\begin{abstract} 
  In a previous paper (Krumm and Vogel 2018 Phys. Rev. A \textbf{97}, 043806) we presented a method to solve the nonlinear Jaynes-Cummings dynamics, describing the quantized motion of a trapped ion exactly, including detuning.
  Here we investigate this model with respect to nonclassical effects, such as squeezing and sub-Poisson statistics.
  We show that for the versatile model under study there exist quantum phenomena beyond squeezing and sub-Poisson statistics, such as anomalous quantum correlations of two non-commuting observables. 
  In particular, it is shown that for the excitation of the zeroth sideband neither squeezing nor sub-Poisson statistics do occur, but anomalous correlations can be verified. 
  Furthermore, it is shown how these anomalous correlation functions can be derived from measured data.
\end{abstract}

\date{\today}
\maketitle

\section{Introduction}
\label{Sec:Introduction}
	In the wide-ranging field of quantum optics, vital areas of interest are the identification, characterization and quantification of nonclassical effects---i.e. effects that can not be explained within Maxwell's theory of classical electrodynamics.
	During the last decades significant efforts were made to develop techniques that allow not only for the theoretical description but also for the experimental verification of nonclassical states.
	Prominent examples are photon antibunching~\cite{Lit:Walls76,Lit:Mandel76,Lit:KDM77}, squeezing~\cite{Lit:W83,Lit:Slusher,Lit:Wu,Lit:Va08,Lit:Va16}, sub-Poisson statistics~\cite{Lit:M79,Lit:S83,Lit:T85}, and entanglement~\cite{Lit:EPR35,Lit:S35,Lit:L00,Lit:B02,Lit:B05,Lit:H09}. 

	On a general basis, nonclassicality can be subdivided into two sets, namely single-time and multi-time nonclassicality.
	This means, there exist effects that can be characterized by using a single point in time and effects which need, for its description, two or more points in time.
	One example of the latter is photon antibunching as two points in time are required for its general analysis.
	
	A general treatment of quantum correlations of radiation fields was introduced in~\cite{Lit:V08}. Based on  normal- and time-ordered correlation functions, it was shown that a plethora of multi-time nonclassicality criteria can be derived. 
	They verify nonclassical effects beyond photon antibunching and, in the special single-time scenario, nonclassicality beyond squeezing and sub-Poisson light.
	Those phenomena include so called anomalous correlations of non-commuting observables~\cite{V91}. Such effects have recently been demonstrated to occur for squeezed coherent light, even for phase  values when squeezing does not occur~\cite{Lit:BK17}.
	
	The question arises whether or not a similar behavior can be found in other, more sophisticated physical systems.
	An encouraging approach can be based on the Jaynes-Cummings model~\cite{Lit:JC63,Lit:P63}, which was widely applied in cavity QED, see e.g.,~\cite{Lit:Har13}.
	Using a vibrational rotating-wave approximation, this model also applies to describe the quantized center-of-mass motion of a trapped ion in a Paul trap~\cite{Lit:B92a,Lit:CBZ94}, for related experiments, see~\cite{Lit:wine13}.
	During the years it became feasible to study many nonclassical motional states of the ion~\cite{Lit:C93,Lit:M96,Lit:F96,Lit:MV96,Lit:GS96a,Lit:GS96b,Lit:G97,Lit:MM96,Lit:G96,Lit:WV97}.
	Under more general conditions, the Hamiltonian describing the dynamics of a trapped ion attains the form of a nonlinear Jaynes-Cummings model~\cite{Lit:VF95}. 
	Recently, the latter was extended to include some frequency mismatch, leading to an explicitly time-dependent dynamics~\cite{Lit:K18}.
	It is noteworthy that related approaches can include the counter-rotating terms of the Hamiltonian, which are neglected withing the vibrational rotating-wave approximation.  
	The corresponding framework is referred to as Quantum Rabi Model, which is for example treated in~\cite{MC2000,MC2012,P15,M16,C17,MC2018}.
		
	In the present paper, we study the nonlinear Jaynes-Cummings dynamics to analyze quantum effects in the atomic center-of-mass motion of a trapped ion, with particular emphasis on anomalous quantum correlations. 
	As those correlations are normal-ordered ones, they are not hindered by vacuum fluctuations which typically occur in the presence of losses.
	In addition, as it was demonstrated in a recent quantum-optics experiment, the anomalous quantum correlations are capable of certifying nonclassicality beyond squeezing; see~\cite{Lit:BK17}. The reduction of quantum noise effects by the use of squeezed states is limited to narrow phase intervals, in particular for strong squeezing. As strongly squeezed states can be easily prepared in the center-of-mass motion of trapped ions, see~\cite{Lit:M96}, this opens new applications of squeezing for phase-noise tolerant applications of trapped ions in quantum technology. Thus, we investigate how anomalous quantum correlations behave in the dynamics of trapped ions driven in the resolved sideband regime and how they can be detected. Moreover, we demonstrate that those correlations can be even prepared when squeezing or sub-Poisson statistics do not persist.
	
	The paper is structured as follows. 
	In Sec.~\ref{Sec:JC} we briefly recapitulate the model under study.
	Afterwards, in Sec.~\ref{Sec:ST-Corr} we analyze in some detail nonclassical phenomena. A detailed consideration of the measurement of the correlation functions under study is provided in Sec.~\ref{Sec:meas}. 
	Finally, we give a summary and some conclusions in Sec.~\ref{Sec:Conclusions}.

\section{The nonlinear Jaynes-Cummings model including detuning}
\label{Sec:JC}

	  The time-independent version of the nonlinear Jaynes-Cummings model was introduced in~\cite{Lit:VF95}.
	  In order to study the influence of time ordering and a time-dependent Hamiltonian in general, we extended the model such that a detuning between the monochromatic driving laser and the electronic transitions can be included~\cite{Lit:K18}.
	  To solve the resulting time-dependent Hamiltonian analytically, the driving laser-field was quantized.
	  That is, the amplitude of the laser, $\beta_0$, was replaced by the corresponding Hilbert-space operator $\hat b$, which obeys the eigenvalue equation
	  \begin{align}
	   \hat b |\beta_0 \rangle = \beta_0 | \beta_0 \rangle,
	  \end{align}
	  where $|\beta_0 \rangle $ is a coherent input state of a cavity mode.
	  In the limit of a strong coherent amplitude, $\beta_0 \gg 1$, the classical solutions of the dynamics are recovered~\cite{Lip18}.
	  In this Section we recapitulate the basic equations and the obtained analytic solutions of the interaction problem.
	  The total Hamiltonian to be studied in the Schr\"odinger picture, including the quantized pump field (cavity field), reads as
	 \begin{align}
	 	  \label{Eq:H}
	 &\hat H=\hat H_0 + \hat H_\text{int}, \\
	&\hat H_0=   \hbar \nu \hat a^\dag \hat a  +\hbar \omega_L \hat b^\dag \hat b+ \hbar \omega_{21} \hat A_{22}, \nonumber \\
	 &\hat H_\text{int} = \hbar \kappa \hat A_{21} \hat b \hat f_k(\hat a^\dag \hat a;\eta) \hat a^k  + \text{H.c.}   \nonumber 
	\end{align}
	
	 The first term of $\hat H_0$ describes the free evolution of the vibrational center-of-mass motion, with the vibrational frequency $\nu$.
	 The second term represents the free evolution of the cavity field, with the laser frequency $ \omega_L=\omega_{21}-k \nu + \Delta \omega$.
	 The free evolution of the electronic degrees of freedom of the two-level ion, with the electronic transition frequency $\omega_{21}=\omega_2-\omega_1$, is given by the third term of $\hat H_0$.
	 The operators $\hat a^\dag$  ($\hat a$) create (annihilate) the quanta of the vibrational mode whose energy levels are equidistantly separated by $\nu$.
	 The atomic $|j\rangle \rightarrow |i\rangle$ transitions are described by the atomic flip operators $\hat A_{ij}=|i\rangle \langle j|$ ($i,j=1,2$).
	 In $\hat H_\text{int}$, $\kappa$ is the coupling of the vibronic system and 
	\begin{align}
	 &\hat f_k(\hat n;\eta)= \frac{1}{2} e^{i \Delta \phi -\eta^2/2} \sum_{l=0}^\infty \frac{(i \eta)^{2l+k}}{l! (l+k)!} \hat a^{\dag l} \hat a^l + \text{H.c.}  \nonumber \\
	 &=\frac{1}{2} e^{i \Delta \phi -\eta^2/2}  \sum_{n=0}^\infty |n\rangle \langle n | \frac{(i \eta)^k n!}{(n+k)!} L_n^{(k)} (\eta^2) + \text{H.c.}   \label{Eq:def-f}
	\end{align}
	 describes the mode structure of the driving laser (standing wave) at the operator-valued position of the ion.
	$L_n^{(k)} $ denotes the generalized Laguerre polynomials, $|n\rangle$ the motional number states, $\hat n=\hat a^\dag \hat a$ the corresponding number operator, $\Delta \phi$ defines the position of the trap potential relative to the laser wave, and $\eta$ is the  Lamp-Dicke parameter.
	 
	The physical interpretation of the Hamiltonian in equation~\eqref{Eq:H} is as follows: 
	a cavity photon is absorbed ($\hat b$) and the trapped ion is excited ($\hat A_{21}$). 
	The transitions of the vibrational states ($\hat f_k(\hat a^\dag \hat a;\eta) \hat a^k $) occur according to the chosen laser frequency such that only the $|n\rangle \leftrightarrow |n-k\rangle$ transitions are driven. 
	That is, we assume we operate in a limit that the vibrational sidebands can be resolved very well (resolved sideband regime).
	Note that the electronic de-excitation process is described by the $\text{H.c.}$ term.
	
	The dynamics of the Hamiltonian in equation~\eqref{Eq:H} is described by the time-evolution operator~\cite{Lit:K18}
	 \begin{align}
	\label{Eq:TimeEvolutionOperator}
	& \hat U (t,t_0) = \sum_{\sigma=\pm} \sum_{m,n=0}^\infty e^{-i  \omega_{mn}^\sigma (t-t_0)}|\psi_{mn}^\sigma \rangle \langle \psi_{mn}^\sigma | \nonumber \\
	 &+\sum_{n=0}^\infty e^{-i  \nu n (t-t_0)} |1,0,n\rangle \langle 1,0,n| \nonumber \\
	 & + \sum_{m=0}^\infty \sum_{q=0}^{k-1} e^{-i [ \nu q +  \omega_L(m+1) ](t-t_0) } |1,m+1,q\rangle \langle1,m+1,q|.
	\end{align}
	The eigenstates of the Hamiltonian read as
	\begin{align}
	|\psi^\pm_{mn} \rangle=c^\pm_{mn}( |2,m,n\rangle + \alpha^\pm_{mn} |1,m+1,n+k \rangle ).
	\end{align}
	In $|i,m,n\rangle$, the $i=1,2$ refer to the electronic excitations, $m$ and $n$ are the photon number of the cavity field and the motional excitation of the ion, respectively.
	Furthermore, one finds
	\begin{align}
	 &c^\pm_{mn}=\frac{1}{\sqrt{1+|\alpha^\pm_{mn} |^2}},   \nonumber \\
	 & \alpha_{mn}^\pm =\frac{\Delta \omega \pm \sqrt{\Delta \omega^2 + |\Omega_{mn}|^2 }}{ \Omega_{mn} },   \nonumber \\
	 & \omega_{mn}^\pm=  \frac{1}{2} \{ \Delta \omega ( 2m+1) + \nu ( 2n-2km) + \omega_{21} (2m+2)  \nonumber \\
	   & \pm \sqrt{\Delta \omega^2 + |\Omega_{mn} |^2} \},\nonumber \\
	   & \Omega_{mn} = 2 \kappa \sqrt{m+1} f_k(n;\eta)\sqrt{\frac{(n+k)!}{n!}}, \nonumber \\
	  &f_k(n;\eta)=\langle n | \hat f_k(\hat n ;\eta) | n \rangle.
	\end{align}
	Based on these solutions, general properties of the center-of-mass motion can be described.

\section{Nonclassicality}
\label{Sec:ST-Corr}

	 During the last decades various criteria to identify nonclassicality of different types were derived.
	The most elementary conditions are those for squeezing~\cite{Lit:W83,Lit:Slusher,Lit:Wu,Lit:Va08,Lit:Va16} and sub-Poisson statistics (Mandel Q parameter)~\cite{Lit:M79,Lit:S83,Lit:T85}.
	 In this section we investigate nonclassical properties and their temporal evolutions in the explicitly time-dependent nonlinear Jaynes-Cummings model.
	 In~\cite{Lit:K18}, it was already shown that the vibrational states are clearly nonclassical for the times under study.
	 Here, we discuss this behavior in more detail, especially with respect to the anomalous quantum correlation effects.
	 
	\subsection{Special nonclassical effects}
	 \label{Sec:SingleTimevs}
	 
	 In the following we denote the nonclassicality criteria by $\mathcal C  $.
	 Squeezing is defined through the negativity of the normal-ordered variance of the quadrature operator, ${\hat x(\varphi;\tau)= e^{i \varphi} \hat a(\tau) + e^{-i \varphi}\hat a^\dag(\tau)}$, see~\cite{Lit:VW06}.
	 Thus, if
	\begin{align}
	 \langle : [ \Delta \hat x (\varphi;\tau) ]^2 : \rangle <0
	\end{align}
	for some $\varphi$-interval, with $\Delta \hat x= \hat x - \langle \hat x \rangle$, the state is referred to as a quadrature squeezed state.
	The normal-ordering prescription orders the operators $\hat a$ and $\hat a^\dag$ such that all creation operators $\hat a^\dag$ are placed to the left of the annihilation operators $\hat a$.
	Consequently, we define the criterion for squeezing as
	\begin{align}
	 \label{Eq:SqueezingCondition}
	 \mathcal C_\text{Sq} := \min_{\varphi \in [0,2\pi )} \left\{  \langle : [ \Delta \hat x (\varphi;\tau) ]^2 : \rangle \right\} <0.
	\end{align}
	That is, if $\mathcal C_\text{Sq}   < 0$ then squeezing occurs at time point $ \tau $.
	Beside the condition in equation~\eqref{Eq:SqueezingCondition}, we consider the Mandel $Q$ parameter~\cite{Lit:M79,Lit:S83,Lit:T85}:
	\begin{align}
	\label{Eq:MandelQ}
	 \mathcal C_\text{SP}  :=  Q(\tau) = \frac{\langle : [\Delta \hat n(\tau) ]^2 :  \rangle }{\langle \hat n(\tau) \rangle} <0.
	\end{align}
	In terms of radiation fields a negative Mandel $Q$ parameter certifies a photocounting statistics of sub-Poisson type.
	Such a statistics does not possess a classical analog and, hence, $ \mathcal C_\text{SP} <  0$ verifies nonclassicality.
	We also consider an anomalous quantum-correlation condition, which cannot be fulfilled by classical states~\cite{V91,Lit:V08},
	\begin{align}
	  \label{Eq:AnomalousSingleTime}
	 &\mathcal C_\text{AC} :=\min_{\varphi \in [0,2\pi )} \left\{   \langle : [\Delta \hat n(\tau) ]^2 : \rangle \langle : [\Delta \hat x(\varphi;\tau)]^2 : \rangle \right. \nonumber \\
	 & \left. - | \langle : \Delta \hat x(\varphi;\tau) \Delta \hat n(\tau) : \rangle|^2 \right\} <0.
	\end{align}
	The squeezing condition [equation~\eqref{Eq:SqueezingCondition}] depends solely on the normal-ordered variance of the quadrature operator $\hat x$.
	The sub-Poisson condition [equation~\eqref{Eq:MandelQ}] depends on the normal-ordered version of the variance of the number operator $\hat n$.
	The anomalous quantum-correlation condition in equation~\eqref{Eq:AnomalousSingleTime} contains contributions of both quantities together with their quantum correlations in the last term.
	Recently, it was shown via homodyne cross-correlation measurements~\cite{Lit:V95} that this condition certifies nonclassicality for radiation fields beyond squeezing~\cite{Lit:BK17}.
	
	To calculate all needed quantities, we express the inequalities in terms of the creation and annihilation operators.
	The condition for squeezing [equation~\eqref{Eq:SqueezingCondition}] reads as
	\begin{align}
	 \label{Eq:SqueezingCondition-long}
	  &\langle : [ \Delta \hat x (\varphi;\tau) ]^2 : \rangle  \nonumber \\
	  &= 2 \text{Re}{\left\{ e^{2 i \varphi} \langle \hat a(\tau)^2 \rangle  \right\} }  - 4 \left[ \text{Re}{\left\{  e^{ i \varphi} \langle \hat a(\tau) \rangle\right\} } \right]^2+2 \langle \hat n(\tau) \rangle  . 
	\end{align}
	$\text{Re}{ \{ z \} } $ denotes the real part of the variable $z$.
	The anomalous correlation function in equation~\eqref{Eq:AnomalousSingleTime} may be rewritten as
	\begin{align}
	& \langle : \Delta \hat x(\varphi;\tau) \Delta \hat n(\tau) : \rangle \nonumber \\
	 &= 2 \text{Re}{\left\{ e^{ i \varphi} \langle \hat n(\tau)  \hat a(\tau) \rangle \right\} } -  2 \text{Re}{ \left\{ e^{ i \varphi} \langle \hat a(\tau) \rangle \right\} }\langle \hat n(\tau) \rangle
	\end{align}
	and
	\begin{align}
	 \langle :  [\Delta \hat n(\tau)]^2 : \rangle &=  \langle : \hat n(\tau)^2 :  \rangle - \langle \hat n(\tau)\rangle^2,
	\end{align}
	with $: \hat n(\tau)^2 :  = \hat n(\tau)^2-\hat n(\tau)$.
	Thus, the sub-Poisson-condition [equation~\eqref{Eq:MandelQ}] can be rewritten as
	\begin{align}
	 \mathcal C_\text{SP} = \frac{\langle \hat n(\tau)^2 \rangle - \langle \hat n(\tau) \rangle^2}{\langle \hat n(\tau) \rangle}-1,
	\end{align}
	which equals the commonly used form of the Mandel Q parameter.

	Since we consider only single-time expectation values of operators that are initially attributed to the motion of the ion, in general denotes by $\hat A(0)$, the expectation values can be calculated via 
	\begin{align}
	 \langle \hat A(t) \rangle = \text{Tr} \left\{ \hat \rho_\text{mot}(t) \hat A(0) \right\},
	\end{align}
	with the reduced motional density matrix $\hat \rho_\text{mot}(t)$. 
	The reduced density matrix of the motional subsystem is obtained by the trace over electronic degrees of freedom and the cavity field,
	  \begin{align}
	  \label{Eq:reddensityMatrix}
	  &\hat \rho_\text{mot}( t,t_0 ) = \sum_{i=1,2} \sum_{m=0}^\infty  \langle i,m | \hat \rho(t,t_0) |i,m \rangle  \nonumber \\
	  &  =  \sum_{m=0}^\infty \frac{|\beta_0|^{2m}e^{-|\beta_0|^2}}{m!}  \sum_{n,n'=0}^\infty  \rho_{n,n'} \nonumber \\
	  & \times  \sum_{\sigma,\sigma'=\pm}   e^{i [ \omega_{mn'}^{\sigma'} -  \omega_{mn}^\sigma]  (t-t_0) } 
	  |c_{mn}^\sigma c_{mn'}^{\sigma'}|^2 \nonumber \\
	  &\times \left\{ |n \rangle \langle n'| + \alpha_{mn}^\sigma ( \alpha_{mn'}^{\sigma'} )^\ast |n+k \rangle \langle n'+k|\right\}.
	 \end{align}
	Here, we introduced ${\rho_{n,n'} = \langle n | \hat \rho_\text{mot}(0) | n' \rangle }$ as the motional input state, the cavity field is in a coherent state
	$|\beta_0 \rangle$, and the electronic degree of freedom of the ion is prepared in the excited state as $|2\rangle$.

\subsection{Analytical results}
 \label{Sec:Results}
	 
	 We recapitulate: nonclassicality at time $\tau $ is certified via $\mathcal C_x  <0$ for $x=\{ \text{Sq},\text{SP}, \text{AC} \}$ (squeezing, sub-Poisson statistics, anomalous quantum correlations), according to 
	 Eqs~\eqref{Eq:SqueezingCondition},~\eqref{Eq:MandelQ}, and~\eqref{Eq:AnomalousSingleTime}.
	 Let us first consider the case where the ion is driven quasi-resonantly to the zeroth sideband.
	 That is, in our model we choose $k=0$, which means that we consider the $|1,n\rangle \leftrightarrow |2,n\rangle$ transitions.
	 
	 \begin{figure*}[t]
	\centering
	\includegraphics*[width=8.6cm]{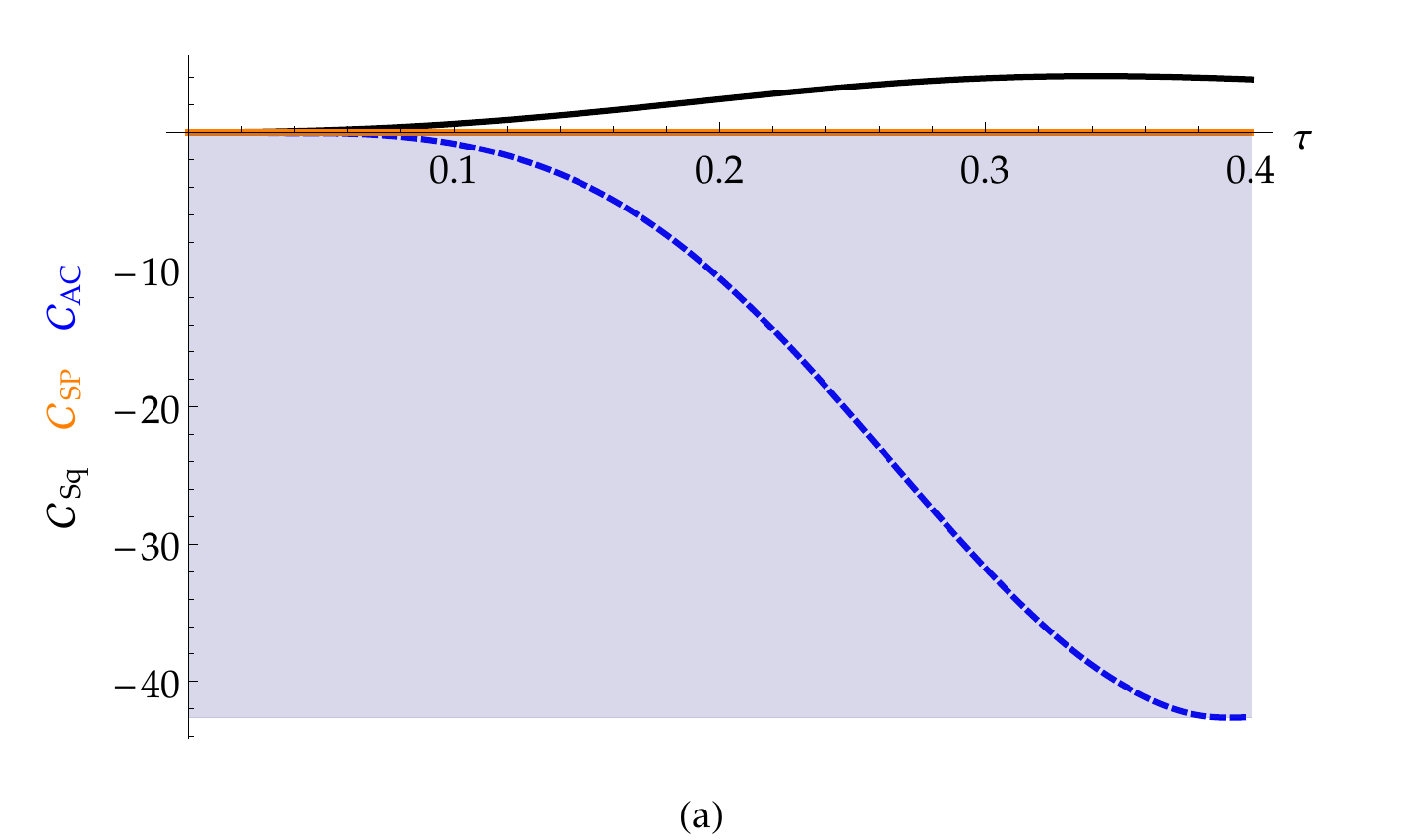}
	\includegraphics*[width=8.6cm]{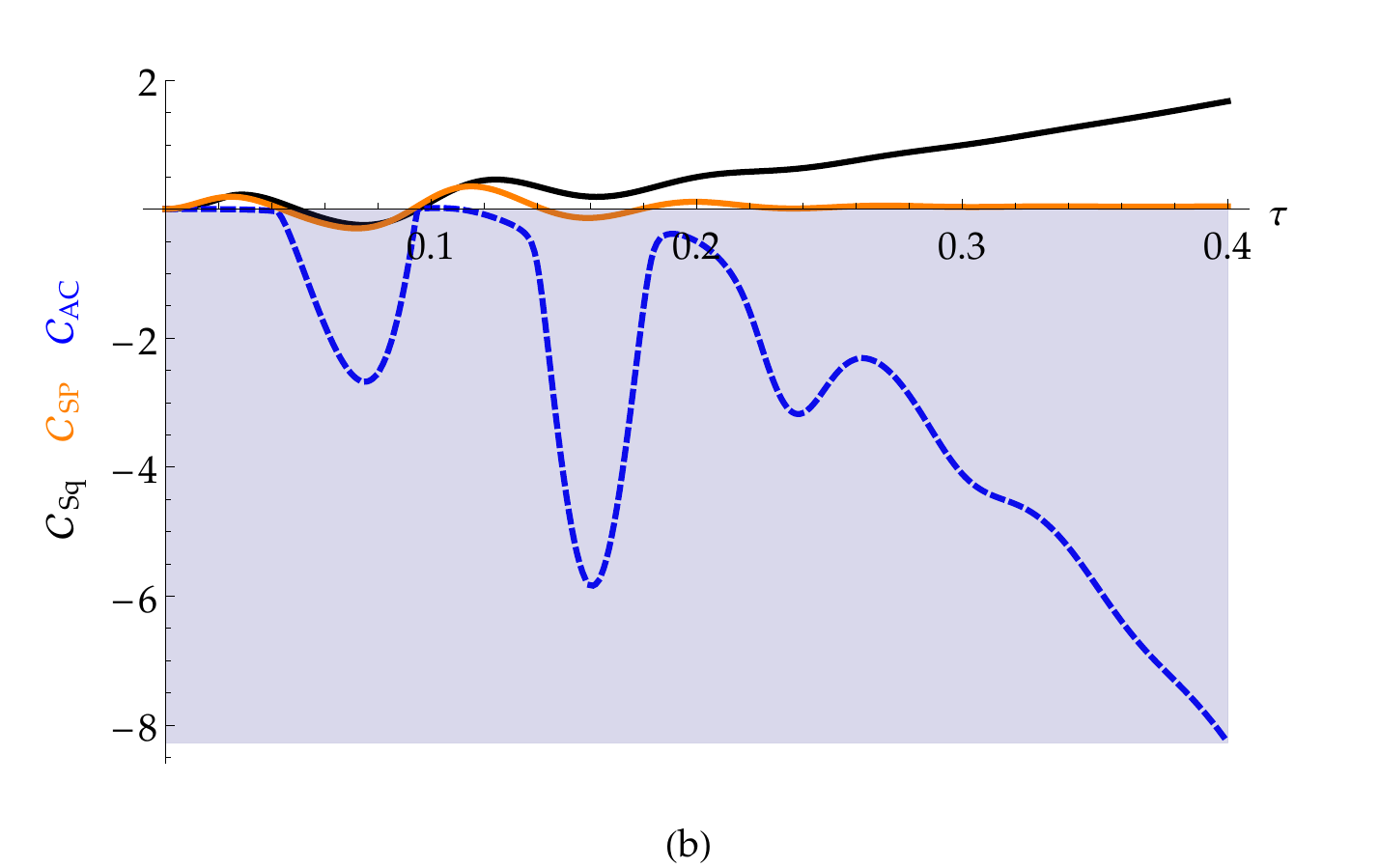}
	\caption{
	  Different nonclassicality criteria are shown as defined in equations~\eqref{Eq:SqueezingCondition} ($\mathcal C_\text{Sq}$; solid black line),~\eqref{Eq:MandelQ} ($\mathcal C_\text{SP}$; solid orange line), 
	  and~\eqref{Eq:AnomalousSingleTime} ($\mathcal C_\text{AC}$; dashed blue line) for excitation to the zeroth sideband $k=0$ (a) and to the second sideband $k=2$ (b).
	  Negative values (shaded area) certify nonclassicality. The motional input state is a coherent state $|\alpha_0 \rangle$.
	 Note that in the case of $\mathcal C_\text{Sq}$ and $\mathcal C_\text{AC}$ the phase $\varphi$ is optimized for each criterion separately.
	  Parameters: $\alpha_0=\sqrt{8}$, $\eta=0.3$, $\Delta \omega/|\kappa|=20$, $\Delta \Phi=0$, $\nu/|\kappa|=5000$, and $\beta_0=100$.
	}\label{Fig:figST1}
	\end{figure*}
	
	According to~\cite{Lit:FV96}, where the authors considered the semiclassical, $|\beta_0| \gg 1$, case with exact resonance, $\Delta \omega=0$, the quantum nondemolition measurement of the motional energy of the trapped ion was proposed. 
	This case, extended by a detuning and a quantized pump, is studied in the following with respect to its nonclassical properties. 
	The pump is chosen to be strong such that the dynamics is close to the semiclassical one, as it was treated in~\cite{Lit:FV96} for the resonant case.
	
	We consider moderate detuning, $\Delta \omega/|\kappa|=20$, and a Lamb-Dicke parameter of $\eta=0.3$.
	 The results are depicted in Fig.~\ref{Fig:figST1} (a).
	Obviously, since $\mathcal C_\text{SP} \geq 0$ and $\mathcal C_\text{Sq} \geq 0$, neither sub-Poisson statistics nor quadrature squeezing are observed.
	Nonclassicality is only revealed by the anomalous correlations as defined in equation~\eqref{Eq:AnomalousSingleTime}.
	The same results are found if other values of $\eta$ or $\Delta \omega$ are considered.
	On the investigated time scales we can, using the considered nonclassicality criteria, certify nonclassicality criteria only through anomalous correlations.
	
	For the same choice of $\eta=0.3$ and $\Delta \omega/ |\kappa|=20$, we now consider the excitation to the second vibrational sideband, $k=2$.
	The results are given in Fig.~\ref{Fig:figST1} (b).
	Naively, one would expect a significant squeezing contribution as the squeezing operator consists of quadratic contributions of the creation and annihilation operators.
	Unexpectedly, we only find small regions where the system is nonclassical regarding the squeezing condition [equation~\eqref{Eq:SqueezingCondition}] and the sub-Poisson condition [equation~\eqref{Eq:MandelQ}].
	Again, the anomalous correlations certify nonclassicality in a very pronounced manner, over nearly the whole considered time range.
	Additionally, for large times we see that the criteria develop in different directions.
	This counterintuitive behavior is caused by the nonlinearities, occurring beyond the Lamb-Dicke regime, which have a significant impact on the dynamics.
	Note that a larger detuning leads to a decrease of the overall strength of the effects.
	
	For convenience, let us visualize the motional state in the corresponding phase-space picture.
	That is, we need to choose an appropriate phase-space distribution.
	Here, we use the regularized version of the Glauber-Sudarshan $P$~function. 
	The $P$~function~\cite{Lit:S63,Lit:G63} itself can be used to express the density operator of an arbitrary state as a pseudo-mixture of coherent states, namely
	\begin{align}
		\hat \rho(t) = \int d^2\alpha P(\alpha;t) |\alpha \rangle \langle \alpha |,
	\end{align}
	where $P(\alpha;t)$ can become negative and even strongly singular.
	A state is referred to as classical state if the corresponding $P$~function has the properties of a classical probability density---i.e., it is non-negative~\cite{Lit:TG65,Lit:M86}.
	However, for the most states the $P$~function is not experimentally accessible due to its singularities.
	Hence, a regularization procedure was introduced in~\cite{Lit:KV10} to transform the ordinary $P$~function into a well behaved phase-space representation, $P_\Omega$, of the state under study.
	This procedure works as follows:
	Since the singularities of $P$ are caused by an unbounded characteristic function $\Phi$, one introduces a suitable filter function $\Omega_w$, with a width $w$. It is constructed such that the filtered function $\Phi_\Omega= \Omega_w\Phi$ is square-integrable for any quantum state.
	Since we are mainly interested in the negativities of $P$, it is important that the filter function must not introduce additional negativities in the filtered $P$~function denoted by $P_\Omega$.
	Thus, the Fourier transform of $\Omega_w$ must be non-negative.
	
\begin{figure}[h]
	\centering
	\includegraphics*[width=8.6cm]{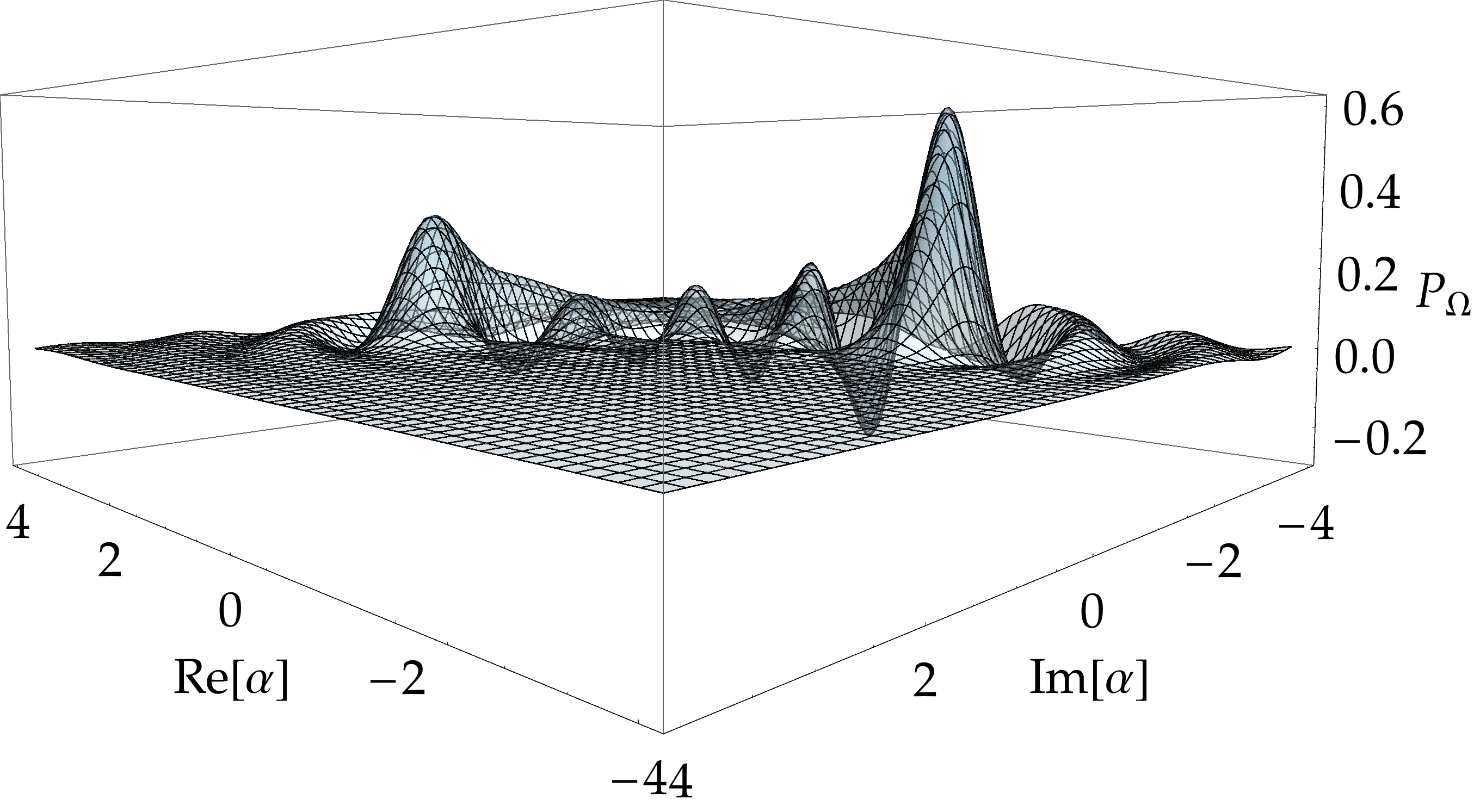}
	\caption{
	  The regularized Glauber-Sudarshan $P$~function, $P_\Omega$, for the excitation to the zeroth sideband at $|\kappa|t=0.2$.
	  The other parameters are equal to those used in Fig.~\ref{Fig:figST1}.
	}\label{Fig:figPST}
	\end{figure}

	Using the procedure outlined in~\cite{Lit:K18}, one may calculate $P_\Omega$ directly out of the reduced density matrix in equation~\eqref{Eq:reddensityMatrix}.
	A plot of $P_\Omega$ is given in Fig.~\ref{Fig:figPST}, where we considered the same situation as in Fig.~\ref{Fig:figST1} (a) for $|\kappa|t=0.2$.
	The depicted state does neither reveal squeezing nor sub-Poisson statistics but only anomalous quantum correlations [see Fig.~\ref{Fig:figST1} (a)].
	The nonclassicality is uncovered by the negative values of $P_\Omega$.
	
	Altogether, we see that for many situations the most commonly used definitions of nonclassicality, such as the negative Mandel Q parameter (sub-Poisson statistics) [equation~\eqref{Eq:MandelQ}] and quadrature squeezing [equation~\eqref{Eq:SqueezingCondition}], fail to certify nonclassicality in the detuned 
	nonlinear Jaynes Cummings model.
	In the scenario $k=0$, where the zeroth sideband is only excited, the anomalous quantum-correlation condition reveals the nonclassical character of the dynamics.
	In the $k=2$ case, the applicability of the criteria, for the purpose to uncover nonclassicality, depends on the choice of $\eta$ and $\Delta \omega$.
	However, the anomalous quantum-correlation condition~\eqref{Eq:AnomalousSingleTime} is a powerful tool to certify nonclassicality for nearly the full timescale under study.
	This underlines the strength of this condition and it encourages one to investigate quantum effects beyond the mostly considered criteria.
	Especially, the excitation to the zeroth sideband, which only reveals its nonclassical character in terms of anomalous correlations, is a promising scenario to further analyze the physical relevance of such quantum signatures.

\section{Measurement}
\label{Sec:meas}
	
\begin{figure}[h]
	\centering
	\includegraphics*[width=8.6cm]{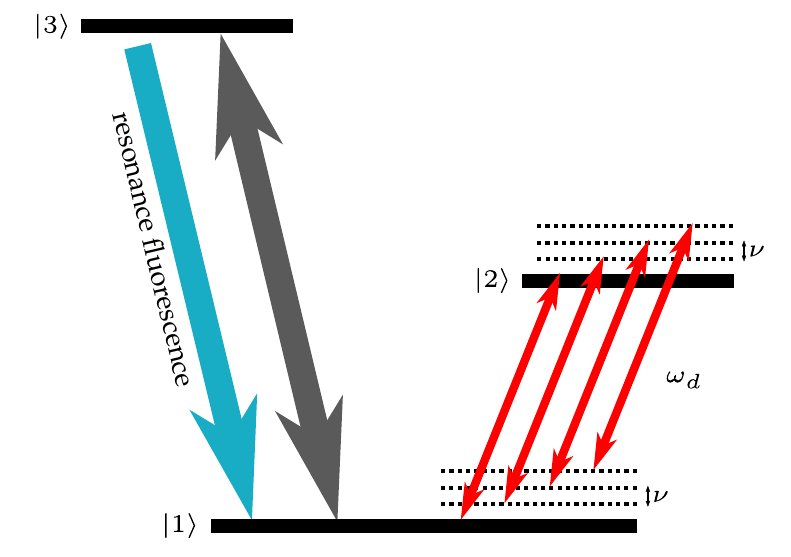}
	\caption{
	 Measurement scheme (reprinted with permission from~\cite{Lit:W97}, copyright (1997) by the American Physical Society) to reconstruct the quantum state of a trapped ion whose vibronic states may be entangled.
	 The strong $|1\rangle \leftrightarrow |3 \rangle $ transition is used to probe the ion's ground state occupation probability.
	  The weak $|1\rangle \leftrightarrow |2 \rangle $ transition leads to an interaction Hamiltonian specified in equation~\eqref{Eq:measH}.
	  The driving laser (red arrows) is detuned to the zeroth sideband: $\omega_d=\omega_{21} $.
	}\label{Fig:measscheme}
	\end{figure}

	In the following we consider the possible measurement of the correlations studied for the quantized motion of the trapped ion.
	For radiation fields, the anomalous correlations were measured recently~\cite{Lit:BK17}.
	However, the reconstruction of the motional state of a trapped ion is a sophisticated problem itself~\cite{Lit:WV95,Lit:WV96}.
	In the following, we are interested in the measurement of the full vibronic quantum state by the technique introduced in~\cite{Lit:W97}, for the purpose to detect entanglement of the vibronic degrees of freedom, see the scheme in Fig.~\ref{Fig:measscheme}.
	
	The strategy is as follows:
	The weak $|1\rangle \leftrightarrow |2 \rangle$ transition is the one whose joint quantum state we are interested in.
	The electronic state is tested by a strong $|1\rangle \leftrightarrow |3\rangle$ transition via the appearance of resonance fluorescence~\cite{Lit:N86,Lit:S86,Lit:B86}.
	If the latter is detected, the ion is in the state $|1\rangle$, otherwise in the state $|2\rangle$.
	The incident laser is tuned to the zeroth sideband, which, in the resolved sideband limit, leads to the interaction Hamiltonian, see~\cite{Lit:FV96},
	\begin{align}
	  \label{Eq:measH}
	 \hat{\mathcal H}_\text{int}= \hbar |\kappa'| \hat f_0 (\hat n;\eta) \hat A_{12} + \text{H.c.}
	\end{align}
	Here we use the notation $ \hat{\mathcal H}_\text{int}$ to distinguish this Hamiltonian from the previously discussed one in equation~\eqref{Eq:H}.
	The operator-valued function $\hat f_0 (\hat n;\eta)$ is defined in equation~\eqref{Eq:def-f}.
	The corresponding time-evolution operator is obtained via $  \hat{\mathcal U}_\text{int}(\tau) = \exp \left( - \frac{i}{\hbar}\hat{\mathcal H}_\text{int} \tau \right)$.
	
	Usually, one focuses on the no-fluorescence events since the motional state is then not disturbed due to recoil effects.
	The initial \textit{probe cycle} is performed as follows:
	First, the motional state is coherently displaced by the amplitude $\alpha$, which can be accomplished via the application of a radio-frequency field.
	Second, the driving laser with frequency $\omega_d$ [Fig.~\ref{Fig:measscheme}] is switched on for a certain interaction time $\tau_1$.
	Afterwards, the electronic state is measured via probing for resonance fluorescence.
	After such a probe cycle, if no fluorescence is detected, the unnormalized density operator of the ion reads as
	\begin{align}
		\hat \rho^{(1)}(\tau_1) = |2 \rangle \langle 2| \otimes \hat \rho_\text{red}^{(1)}(\tau_1),
	\end{align}
	with $\hat \rho_\text{red}^{(1)}(\tau)=\langle 2 |  \hat{\mathcal U}_\text{int}(\tau_1)   \hat \rho (\alpha)\hat{\mathcal U}^\dag_\text{int}(\tau_1)   | 2 \rangle$ and
	$\hat \rho (\alpha)= \hat D^\dag(\alpha) \hat \rho(0) \hat D(\alpha) $, where $\hat D(\beta) = \exp ( \beta \hat a^\dag - \beta^\ast \hat a) $ is the coherent displacement operator.
	Here, $\hat \rho(0)$ denotes the density operator of the vibronic degrees of freedom.
	This means that the cavity mode is traced out.
	As long as the coherent amplitude $\beta_0$ is sufficiently large---i.e., the dynamics is close to the semiclassical case---$\hat \rho(0)$ contains the complete information of the nonclassical properties.
	
	Applying $K$ of those probe cycles, with interaction times $\tau_1,\dots,\tau_K$, yields the diagonal elements (still unnormalized)
	\begin{align}
		&\langle n | \hat \rho^{(K)}(\tau_K) | n \rangle  \nonumber \\
		&= \prod_{q=2}^K \cos^2 \left(  |\kappa'| L_n(\eta^2) e^{-\eta^2/2} \tau_q \right)  \langle n | \hat \rho^{(1)} (\tau_1) | n \rangle.
	\end{align}
	The probability to obtain such a sequence of cycles, equals the trace of the latter expression.
	For appropriately chosen interaction times, see the end of Sec. III of~\cite{Lit:W97}, this probability can directly be related to the displaced density operator elements $\rho_{ij}^{nn} (\alpha) \equiv \langle i,n| \hat \rho(\alpha) |j,n \rangle$ for $i,j=1,2$.
	Using these elements, $\rho_{ij}^{nn} (\alpha)$, one can derive the Wigner-function matrix straightforwardly~\cite{Lit:W97}:
	\begin{align}
		\label{Eq:WignerMatrix}
		W_{ij}(\alpha) = \frac{2}{\pi} \sum_{n=0}^\infty (-1)^n \rho_{ij}^{nn} (\alpha) .
	\end{align}
	The latter is a unification of the ordinary Wigner function~\cite{Lit:W32}, including the electronic degrees of freedom.
	Hence it was shown that using $W_{ij}(\alpha) $ one can uncover entanglement between the motional and electronic states of the ion which would not be verified by using only the reduced density matrix.
	However, nowadays we have experimental access to the regularized $P$~function which possesses several advantages over other quasiprobabilites~\cite{Lit:F12,Lit:A13}.
	Remarkably, in~\cite{Lit:A17} a regularized hybrid version of the $P$~function was introduced, unifying the description of continuous- and discrete-variable systems.
	The definition applies here as well and reads as
	\begin{align}
		\label{Eq:hybridP}
		P_{ij}(\alpha)= \langle \hat  A_{ji} \otimes : \hat \delta (\hat a-\alpha ) : \rangle ,
	\end{align}
	with $\hat A_{ji}=|j\rangle \langle i|$.
	Using this definition, one can define the overall density operator of the system as 
	\begin{align}
		\hat \rho = \sum_{ij} \int d^2\alpha P_{ij}(\alpha) |i\rangle \langle j| \otimes|\alpha \rangle \langle \alpha | .
	\end{align}
	The questions arise how the anomalous moments can be obtained and how $P_{ij}(\alpha)$ can be reconstructed by using the scheme in Fig.~\ref{Fig:measscheme}.
	
	Let us start with the definition given in equation~\eqref{Eq:WignerMatrix}.
	Applying the inverse Fourier transform yields the characteristic-function matrix of the Wigner-function matrix (indicated by the index $\text{W}$),
	\begin{align}
		\label{Eq:WignermatrixCF}
		\Phi_{ij,\text{W}}(\beta) =  \int d^2\alpha W_{ij}(\alpha) e^{\beta \alpha^\ast - \beta^\ast \alpha} = \langle \hat A_{ji} \otimes \hat D(\beta) \rangle.
	\end{align}
	To transform the symmetric ordered function $\Phi_{ij,\text{W}}(\beta) $ into normal order, one may apply the Baker-Campbell-Hausdorff formula to obtain the Fourier transform of the $P$-function matrix $P_{ij}(\alpha)$,
	\begin{align}
		\label{Eq:CFmatrix}
		\Phi_{ij}(\beta)= \langle \hat A_{ji} \otimes :\hat D(\beta) : \rangle =  \langle \hat A_{ji} \otimes \hat D(\beta) \rangle e^{|\beta|^2/2} .
	\end{align}
	As soon as the latter function is derived from $W_{ij}(\alpha)$, one may calculate its trace over the electronic degrees of freedom to obtain the characteristic function of the motional subsystem:
	\begin{align}
		\sum_{i=1}^2 \Phi_{ii}(\beta) = \langle :\hat D(\beta) : \rangle .
	\end{align}
	Differentiation yields the expectation values needed in equation~\eqref{Eq:AnomalousSingleTime}.
	In principle, all possible combinations of normal-ordered moments can be obtained in this way.
	For example, in 
	\begin{align}
	 \langle : \Delta \hat x(\varphi) \Delta \hat n(\tau) : \rangle = \langle : \hat x(\varphi) \hat n : \rangle - \langle \hat x (\varphi)\rangle \langle \hat n\rangle
	\end{align}
	 one may derive all terms as follows, defining $\beta=|\beta|e^{i \varphi_\beta}$:
	 \begin{align}
	 \label{Eq:anomomentsexp}
	 	&\langle \hat x(\varphi) \rangle = \frac{1}{i} \frac{\partial}{\partial |\beta| }  \langle :\hat D(\beta) : \rangle \Big|_{|\beta|=0, \varphi_\beta=\varphi+\pi/2}, \nonumber \\
	 	& \langle \hat n\rangle = - \frac{\partial}{\partial \beta } \frac{\partial}{\partial \beta^\ast}   \langle :\hat D(\beta) : \rangle \Big|_{|\beta|=0}, \nonumber \\
	 	& \langle : \hat x(\varphi) \hat n : \rangle = \frac{-1}{i} \frac{\partial}{\partial \beta } \frac{\partial}{\partial \beta^\ast}   \frac{\partial}{\partial |\beta| }  \langle :\hat D(\beta) : \rangle \Big|_{|\beta|=0, \varphi_\beta=\varphi+\pi/2}.
	 \end{align}
	Via $\frac{\partial}{\partial |\beta|} = \left( e^{i \varphi_\beta} \frac{\partial}{\partial \beta} + e^{-i \varphi_\beta} \frac{\partial}{\partial \beta^\ast} \right)$ all moments can be derived with respect to the derivatives of $\beta$ and $\beta^\ast$.
	
	Furthermore, we may calculate the regularized $P$-function matrix out of equation~\eqref{Eq:CFmatrix} via multiplication of an appropriate filter function $\Omega_w(\beta)$~\cite{Lit:KV10,Lit:BK14} (with a width $w$).
	A subsequent Fourier transformation and the usage of equation~\eqref{Eq:WignermatrixCF} yields
	\begin{align}
		&P_{ij,\Omega}(\alpha) = \frac{1}{\pi^2}  \int d^2\beta \, \Omega_w(\beta) e^{\alpha \beta^\ast - \alpha^\ast \beta} \Phi_{ij}(\beta)  \nonumber \\
		&=   \int \frac{ d^2\alpha'}{\pi^2} W_{ij}(\alpha') \underbrace{ \int d^2\beta \, \Omega_w(\beta)e^{|\beta|^2/2} e^{ \beta^\ast (\alpha-\alpha') -  \beta (\alpha^\ast -\alpha'^\ast)}   }_{:=\Lambda_\Omega(\alpha,\alpha')}.
	\end{align}
	Hence, using equation~\eqref{Eq:WignerMatrix} we finally arrive at
	\begin{align}
		\label{Eq:reconstrP}
		&P_{ij,\Omega}(\alpha) =\frac{2}{\pi^3}  \sum_{n=0}^\infty (-1)^n   \int d^2\alpha' \Lambda_\Omega(\alpha,\alpha')  \rho_{ij}^{nn} (\alpha') ,
	\end{align}
	where $\rho_{ij}^{nn} (\alpha) \equiv \langle i,n| \hat \rho(\alpha) |j,n \rangle$ for $i,j=1,2$.
	Thus, out of the Wigner-function matrix we derive the moments in equation~\eqref{Eq:anomomentsexp} and furthermore, out of the $\rho_{ij}^{nn}  (\alpha)$, we may obtain the regularized $P$-function matrix.
	The trace $\sum_{i=1}^2 P_{ii,\Omega}(\alpha)$  would yield the regularized $P$ representation including merely the motional subsystem, which we discussed for a special case in Fig.~\ref{Fig:figPST}. 
	
	Note that  in equation~\eqref{Eq:reconstrP} one needs to evaluate an integral over the whole $\alpha'$-plane.
	One can avoid this integration by using an alternative approach related to the ideas presented in~\cite{Lit:KieselWitness,Lit:BK16}.
	The \textit{nonclassicality witnesses} for harmonic oscillators, which also apply here, lead to the expression
	\begin{align}
		\label{Eq:witness}
		&P_{ij,\Omega}(\alpha) \nonumber \\
		&=\frac{w^2}{16} \sum_{m=0}^\infty \frac{(-w^2/4)^m}{[(m+1)!]^2} \binom{2m+2}{m} \langle \hat A_{ji} \otimes : \hat n(\alpha)^m : \rangle,
	\end{align}
	where $\hat n(\alpha) = \hat D(\alpha) \hat n \hat D^\dag(\alpha)$.
	This result, expressed in terms of normal-ordered displaced-number moments, is obtained via the application of a particular disc-function filter.
	Inserting the expression
	\begin{align}
	& \langle \hat A_{ji} \otimes : \hat n(\alpha)^m : \rangle =  \sum_{n=m}^\infty  \rho_{ij}^{nn} (\alpha)  \frac{n!}{(n-m)!}
	\end{align}
	in equation~\eqref{Eq:witness}, we directly relate the regularized $P$-function matrix, $P_{ij,\Omega}(\alpha)$, to
	the elements $ \rho_{ij}^{nn} (\alpha)$.
	Especially, we in this formulation we do not have to evaluate an integral over the complex plane, as in equation~\eqref{Eq:reconstrP}.
	We can choose a certain value $\alpha$ and calculate $P_{ij,\Omega}$ at this point in phase-space.
	
	\section{Summary and Conclusions}
	\label{Sec:Conclusions}
	
	In this work we studied nonclassical properties of the recently introduced generalization of the nonlinear Jaynes-Cummings model for the vibronic dynamics of a trapped ion---including a quantized pump field and a small detuning with respect to the vibronic excitation in the resolved sideband regime.
	We showed that for the excitation of the zeroth and second sideband the so-called anomalous quantum correlations of non-commuting observables certify nonclassicality when established criteria, verifying sub-Poisson number statistics or quadrature squeezing, fail.
	Especially, in the case of driving the zeroth sideband, the anomalous quantum-correlation condition is the favored one that uncovers the nonclassicality.
	In addition, we studied the influence of the nonlinearities occurring beyond the Lamb-Dicke regime as well as the detuning from resonance.
	The great importance of the anomalous quantum correlations in the dynamics under study raises the question whether these phenomena may be useful for practical applications in quantum technologies.
	In any case, the verification of the nonclassical nature of the system under study through anomalous quantum correlations is of fundamental interest---in particular if standard quantum signatures (e.g, squeezing and sub-Poisson statistics) are negligibly small.
	
	To access the studied quantum signatures in experiments, we studied the possibilities to determine the needed correlations from measured data. For this aim, a measurement technique is suited which was originally proposed for the purpose to verify entanglement within the vibronic quantum system of the trapped ion.  
          We show in detail how the needed moments and correlation functions, including those characterizing  the anomalous quantum correlations, are obtained from measured quantities. 
          In the underlying measurement scenario, the Wigner-function matrix was considered as the quantity to be determined. In the present paper we demonstrated how the regularized version of the Glauber-Sudarshan $P$-function matrix can be obtained from the Wigner-function matrix. 
          This is needed as the desired correlation functions for analyzing the quantum effects of interest are normal-ordered ones. 
           The advantage of normal ordering consists in the fact that these correlations are robust against losses and they are not washed out by vacuum fluctuations which are caused by losses.
	Based on these techniques, very general quantum effects in the vibronic degrees of freedom of trapped ions may be studied.


\begin{thebibliography}{100}

	\bibitem{Lit:Walls76}
		H. J. Carmichael and D. F. Walls,
		A quantum-mechanical master equation treatment of the dynamical Stark effect,
		J. Phys. B \textbf{9}, 1199 (1976).
	\bibitem{Lit:Mandel76}
		H. J. Kimble and L. Mandel,
		Theory of resonance fluorescence,
		Phys. Rev. A \textbf{13}, 2123 (1976).
	\bibitem{Lit:KDM77}
		H. J. Kimble, M. Dagenais, and L. Mandel,
		Photon Antibunching in Resonance Fluorescence,
		Phys. Rev. Lett. \textbf{39}, 691 (1977).
	\bibitem{Lit:W83}
		D. F. Walls,
		Squeezed states of light,
		Nature (London) \textbf{306}, 141 (1983).
        \bibitem{Lit:Slusher}
		R. E. Slusher, L. W. Hollberg, B. Yurke, J. C. Mertz, and J. F. Valley,
                Observation of Squeezed States Generated by Four-Wave Mixing in an Optical Cavity,
                Phys. Rev. Lett. \textbf{55}, 2409 (1985).
        \bibitem{Lit:Wu}
		L.-A. Wu, H. J. Kimble, J. L. Hall, and H. Wu,
		Generation of Squeezed States by Parametric Down Conversion,
                Phys. Rev. Lett. \textbf{57}, 2520 (1986).
        \bibitem{Lit:Va08}
		H. Vahlbruch, M. Mehmet, S. Chelkowski, B. Hage, A. Franzen, N. Lastzka, S. Go\ss{}ler, K. Danzmann, and R. Schnabel,
		Observation of Squeezed Light with 10-db Quantum-Noise Reduction,
		Phys. Rev. Lett. \textbf{100}, 033602 (2008).
        \bibitem{Lit:Va16}
		H. Vahlbruch, M. Mehmet, K. Danzmann, and R. Schnabel,
		Detection of 15 dB Squeezed States of Light and their Application for the Absolute Calibration of Photoelectric Quantum Efficiency,
		Phys. Rev. Lett. \textbf{117}, 110801 (2016).
         \bibitem{Lit:M79}
		L. Mandel,
		Sub-Poissonian photon statistics in resonance fluorescence, 
		Opt. Lett. \textbf{4}, 205 (1979).
	\bibitem{Lit:S83}
		R. Short and L. Mandel,
		Observation of Sub-Poissonian Photon Statistics,
		Phys. Rev. Lett. \textbf{51}, 384 (1983).
	\bibitem{Lit:T85}
		M. C. Teich and B. E. A. Saleh ,
		Observation of sub-Poisson Franck–Hertz light at 253.7 nm ,
		J. Opt. Soc. Am. B\textbf{2}, 275 (1985).
	\bibitem{Lit:EPR35}
		A. Einstein, N. Rosen, and B. Podolsky,
		Can Quantum-Mechanical Description of Physical reality Be Considered Complete?,
		Phys. Rev. \textbf{47}, 777 (1935).
	\bibitem{Lit:S35}
		E. Schr\"odinger,
		Die gegenw\"artige Situation in der Quantenmechanik,
		Naturwiss. \textbf{23}, 807 (1935).
	\bibitem{Lit:L00}
		 M. Lewenstein, D. Bru\ss, J. I. Cirac, B. Kraus, M. Kus, J. Samsonowicz, A. Sanpera, and R. Tarrach,
		 Separability and distillability in composite quantum systems-a primer,
		 J. Mod. Opt. \textbf{47}, 2481 (2000).
	\bibitem{Lit:B02}
	 	D. Bru\ss,
		 Characterizing entanglement,
		  J. Math. Phys. \textbf{43}, 4237 (2002).
	\bibitem{Lit:B05}
		 N. Brunner, N. Gisin, and V. Scarani,
	 	 Entanglement and non-locality are different resources,
		  New J. Phys. \textbf{7}, 88 (2005).
	\bibitem{Lit:H09}
		R. Horodecki, P. Horodecki, M. Horodecki, and K. Horodecki,
		Quantum entanglement,
		Rev. Mod. Phys. \textbf{81}, 865 (2009).
	\bibitem{Lit:V08}
		W. Vogel,
		Nonclassical Correlation Properties of Radiation Fields,
		Phys. Rev. Lett. \textbf{100}, 013605 (2008).
	\bibitem{V91}
		W. Vogel, 
		Squeezing and anomalous moments in resonance fluorescence,
    		Phys. Rev. Lett. \textbf{67}, 2450 (1991). 
	\bibitem{Lit:BK17}
		B. K\"uhn, W. Vogel, M. Mraz, S. K\"ohnke, and B. Hage,
		Anomalous Quantum Correlations of Squeezed Light,
		Phys. Rev. Lett. \textbf{118}, 153601 (2017).
	\bibitem{Lit:JC63}
		E. T. Jaynes and F. W. Cummings, 
		Comparison of Quantum and Semiclassical Radiation Theories with Application to the Beam Maser,
		Proc. IEEE \textbf{51}, 89 (1963).
	\bibitem{Lit:P63}
		H. Paul,
		Induzierte Emission bei starker Einstrahlung,
		Ann, Phys. (Leipzig) \textbf{11}, 411 (1963).
	\bibitem{Lit:Har13}
		S. Haroche,
		Nobel Lecture: Controlling photons in a box and exploring the quantum to classical boundary,
		Rev. Mod. Phys. \textbf{85}, 1083 (2013).
	\bibitem{Lit:B92a}
		C. A. Blockley, D. F. Walls, and H. Risken, 
		Quantum Collapses and Revivals in a Quantized Trap,
		Europhys. Lett. \textbf{17} , 509 (1992).
	\bibitem{Lit:CBZ94}
		J. I. Cirac, R. Blatt, A. S. Parkins, and P. Zoller,
		Quantum collapse and revival in the motion of a single trapped ion,
		Phys. Rev. A \textbf{49}, 1202 (1994).
           \bibitem{Lit:wine13}
           	D. J. Wineland,
           	Nobel Lecture: Superposition, entanglement, and raising Schr\"odinger’s cat,
           	Rev. Mod. Phys. \textbf{85}, 1103 (2013).
	\bibitem{Lit:C93}
		J. I. Cirac, A. S. Parkins, R. Blatt, and P. Zoller,
		{''Dark'' squeezed states of the motion of a trapped ion},
		Phys. Rev. Lett. \textbf{70}, 556 (1993).
	\bibitem{Lit:M96}
		D. M. Meekhof, C. Monroe, B. E. King, W. M. Itano, and D. J. Wineland,
		{Generation of Nonclassical Motional States of a Trapped Atom},
		Phys. Rev. Lett. \textbf{76}, 1796 (1996).
	\bibitem{Lit:F96}
		R. L. de Matos Filho and W. Vogel,
		{Even and Odd Coherent States of the Motion of a Trapped Ion},
		Phys. Rev. Lett. \textbf{76}, 608 (1996).
	\bibitem{Lit:MV96}
		R. L. de Matos Filho and W. Vogel,
		{Nonlinear Coherent States},
		Phys. Rev. A \textbf{54}, 4560 (1996).
	\bibitem{Lit:GS96a}
		S.-C. Gou, J. Steinbach, and P. L. Knight,
		{Dark pair coherent states of the motion of a trapped ion},
		Phys. Rev. A \textbf{54}, R1014(R) (1996).
	\bibitem{Lit:GS96b}
		S.-C. Gou, J. Steinbach, and P. L. Knight,
		{Vibrational pair cat states},
		Phys. Rev. A \textbf{54}, 4315 (1996).
	\bibitem{Lit:G97}
		C. C. Gerry, S.-C. Gou, and J. Steinbach,
		{Generation of motional SU(1,1) intelligent states of a trapped ion},
		Phys. Rev. A \textbf{55}, 630 (1997).
	\bibitem{Lit:MM96}
		C. Monroe, D. M. Meekhof, B. E. King, and D. J. Wineland,
		{A ''Schr\"odinger Cat'' Superposition State of an Atom},
		Science \textbf{272}, 1131 (1996).
	\bibitem{Lit:G96}
		C. C. Gerry,
		{Generation of Schr\"odinger cats and entangled coherent states in the motion of a trapped ion by a dispersive interaction},
		Phys. Rev. A \textbf{55}, 2478 (1997).
	\bibitem{Lit:WV97}
		S. Wallentowitz and W. Vogel,
		{Quantum-mechanical counterpart of nonlinear optics},
		Phys. Rev. A \textbf{55}, 4438 (1997).
	\bibitem{Lit:VF95}
		W. Vogel and R. L. de Matos Filho, 
		Nonlinear Jaynes-Cummings dynamics of a trapped ion,
		Phys. Rev. A \textbf{52}, 4214 (1995).
	\bibitem{Lit:K18}
		F. Krumm and W. Vogel,
		Time-dependent nonlinear Jaynes-Cummings dynamics of a trapped ion,
		Phys. Rev. A \textbf{97}, 043806 (2018).
	\bibitem{MC2000}
		H. Moya-Cessa and P. Tombesi,
		Filtering number states of the vibrational motion of an ion,
		Phys. Rev. A \textbf{61}, 025401 (2000).
	\bibitem{MC2012} 
		H. Moya-Cessa,  F. Soto-Eguibar, J. M. Vargas-Martinez, R. Juarez-Amaro, and A. Zuniga-Segundo,
		Ion-laser interactions: The most complete solution,
		Phys. Rep. {\bf 513}, 229 (2012). 
	\bibitem{P15}
		J. S. Pedernales, I. Lizuain, S. Felicetti, G. Romero, L. Lamata, and E. Solano,
		Quantum Rabi Model with Trapped Ions,
		Sci. Rep. \textbf{5}, 15472 (2015).
	\bibitem{M16}
		H. M. Moya-Cessa,
		Fast Quantum Rabi Model with Trapped Ions,
		Sci. Rep. \textbf{6}, 38961 (2016).
	\bibitem{C17}
		X.-H. Cheng, I. Arrazola, J. S. Pedernales, L. Lamata, X. Chen, and E. Solano,
		Nonlinear Quantum Rabi Model in Trapped Ions,
		Phys. Rev. A \textbf{97}, 023624 (2018). 
	\bibitem{MC2018}
		J. Casanova, R. Puebla, and H. Moya-Cessa,
		Connecting $n$th order generalised quantum Rabi models: Emergence of nonlinear spin-boson coupling via spin rotations,
		npj Quantum Inf. \textbf{4}, 47 (2018)
	\bibitem{Lip18} 
		T. Lipfert, F. Krumm, M. I. Kolobov, and W. Vogel,
		Time ordering in the classically driven nonlinear Jaynes-Cummings model, 
		Phys. Rev. A \textbf{98}, 06381 (2018).
	\bibitem{Lit:VW06}
		W. Vogel and D.-G. Welsch,
		\textit{Quantum Optics}, 3rd ed.
		(Wiley-VCH, New York, 2006).
	\bibitem{Lit:V95}
		W. Vogel,
		Homodyne correlation measurements with weak local oscillators,
		Phys. Rev. A \textbf{51}, 4160 (1995).
	\bibitem{Lit:FV96}
		R. L. de Matos Filho and W. Vogel,
		Quantum Nondemolition Measurement of the Motional Energy of a Trapped Atom,
		Phys. Rev. Lett. \textbf{76}, 4520 (1996).
	\bibitem{Lit:S63}
		E. C. G.  Sudarshan,
		Equivalence  of  Semiclassical  and Quantum Mechanical Descriptions of Statistical Light Beams, 
		Phys. Rev. Lett. \textbf{10}, 277 (1963).
	\bibitem{Lit:G63}
		R. J. Glauber, 
		Coherent and incoherent states of the radiation field, 
		Phys. Rev. \textbf{131}, 2766 (1963).
	\bibitem{Lit:TG65}
		U. M. Titulaer and R. J. Glauber, 
		Correlation functions for coherent fields,
		Phys. Rev. \textbf{140}, B676 (1965).
	\bibitem{Lit:M86}
		L.  Mandel,  
		Non-classical  states  of  the  electromagnetic field, 
		Phys. Scripta \textbf{T12}, 34 (1986).
	\bibitem{Lit:KV10}
          		 T. Kiesel and W. Vogel,
		Nonclassicality filters and quasi-probabilities,
		Phys. Rev. A \textbf{82}, 032107 (2010). 
	\bibitem{Lit:WV95}
		S. Wallentowitz and W. Vogel,
		Reconstruction of the Quantum Mechanical State of a Trapped Ion,
		Phys. Rev. Lett. \textbf{75}, 2932 (1995).
	\bibitem{Lit:WV96}
		S. Wallentowitz and W. Vogel,
		Motional quantum states of a trapped ion: Measurement and its back action,
		Phys. Rev. A \textbf{54}, 3322 (1996).
	\bibitem{Lit:W97}
		S. Wallentowitz, R. L. de Matos Filho, and W. Vogel,
		Determination of entangled quantum states of a trapped atom,
		Phys. Rev. A \textbf{56}, 1205 (1997).
	\bibitem{Lit:N86}
		W. Nagourney, J. Sandberg, and H. Dehmelt,
		Shelved optical electron amplifier: Observation of quantum jumps,
		Phys. Rev. Lett. \textbf{56}, 2797 (1986).
	\bibitem{Lit:S86}
		Th. Sauter, W. Neuhauser, R. Blatt, and P. E. Toschek,
		Observation of Quantum Jumps,
		Phys. Rev. Lett. \textbf{57}, 1696 (1986).
	\bibitem{Lit:B86}
		J. C. Bergquist, Randall G. Hulet, Wayne M. Itano, and D. J. Wineland,
		Observation of Quantum Jumps in a Single Atom,
		Phys. Rev. Lett. \textbf{57}, 1699 (1986).
	\bibitem{Lit:W32}
		E. Wigner,
		On the Quantum Correction For Thermodynamic Equilibrium,
		Phys. Rev. \textbf{40}, 749 (1932).
	\bibitem{Lit:F12}
		A. Ferraro and M. G. A. Paris,
		Nonclassicality Criteria from Phase-Space Representations and Information-Theoretical Constraints Are Maximally Inequivalent,
		Phys. Rev. Lett. 108, 260403 (2012).
	\bibitem{Lit:A13}
		E. Agudelo, J. Sperling, and W. Vogel,
		Quasiprobabilities for multipartite quantum correlations of light,
		Phys. Rev. A 87, 033811 (2013).
	\bibitem{Lit:A17}
		E. Agudelo, J. Sperling, L. S. Costanzo, M. Bellini, A. Zavatta, and W. Vogel,,
		Conditional Hybrid Nonclassicality,
		Phys. Rev. Lett. \textbf{119}, 120403 (2017).
	\bibitem{Lit:BK14}
		B. K\"uhn and W. Vogel,
		Visualizing nonclassical effects in phase space,
		Phys. Rev. A \textbf{90}, 033821 (2014).
	\bibitem{Lit:KieselWitness}
		T. Kiesel and W. Vogel,
		Universal nonclassicality witnesses for harmonic oscillators,
		Phys. Rev. A \textbf{85}, 062106 (2012).
	\bibitem{Lit:BK16}
		B. K\"uhn and W. Vogel,
		Unbalanced Homodyne Correlation Measurements,
		Phys. Rev. Lett. \textbf{116}, 163603 (2016).
	%%% new literature referee 1
	
	
\end{thebibliography}
\end{document}